\documentclass[a4paper,11pt]{article}
\pdfoutput=1 

\usepackage{jcappub} 

\usepackage[T1]{fontenc} 
\usepackage{booktabs} 
\newcommand{\ra}[1]{\renewcommand{\arraystretch}{#1}} 

\title{\boldmath Angular correlation of cosmic neutrinos with
  ultrahigh-energy cosmic rays and implications for their sources}

\author{Reetanjali Moharana} \author{and Soebur Razzaque}
\affiliation{Department of Physics, University of Johannesburg,\\ P. O.
  Box 524, Auckland Park 2006, South Africa}

\emailAdd{reetanjalim@uj.ac.za}
\emailAdd{srazzaque@uj.ac.za}

\abstract{Cosmic neutrino events detected by the IceCube Neutrino
  Observatory with energy $\gtrsim 30$ TeV have poor angular
  resolutions to reveal their origin. Ultrahigh-energy cosmic rays
  (UHECRs), with better angular resolutions at $>60$ EeV energies, can
  be used to check if the same astrophysical sources are responsible
  for producing both neutrinos and UHECRs. We test this hypothesis,
  with statistical methods which emphasize invariant quantities, by
  using data from the Pierre Auger Observatory, Telescope Array and
  past cosmic-ray experiments. We find that the arrival directions of
  the cosmic neutrinos are correlated with $\ge 100$ EeV UHECR arrival
  directions at confidence level $\approx 90\%$. The strength of the
  correlation decreases with decreasing UHECR energy and no
  correlation exists at energy $\sim 60$ EeV. A search in
  astrophysical databases within $3^\circ$ of the arrival directions
  of UHECRs with energy $\ge 100$ EeV, that are correlated with the
  IceCube cosmic neutrinos, resulted in 18 sources from the {\it
    Swift}-BAT X-ray catalog with redshift $z\le 0.06$. We also found
  3 objects in the K\"uhr catalog of radio sources using the same
  criteria. The sources are dominantly Seyfert galaxies with Cygnus A
  being the most prominent member. We calculate the required neutrino
  and UHECR fluxes to produce the observed correlated events, and
  estimate the corresponding neutrino luminosity (25 TeV--2.2 PeV) and
  cosmic-ray luminosity (500 TeV--180 EeV), assuming the sources are
  the ones we found in the {\it Swift}-BAT and K\"uhr catalogs. We
  compare these luminosities with the X-ray luminosity of the
  corresponding sources and discuss possibilities of accelerating
  protons to $\gtrsim 100$ EeV and produce neutrinos in these
  sources.}

\begin{document}
\maketitle
\flushbottom

\section{Introduction}
\label{sec:intro}

The IceCube Neutrino Observatory, the world's largest neutrino
detector, has recently published neutrino events collected over 3-year
period with energy in the $\sim 30$ TeV--$2$ PeV range
\cite{Aartsen:2014gkd}. Shower events, most likely due to $\nu_e$ or
$\nu_\tau$ charge current $\nu N$ interactions, dominate the event
list (28 including 3 events with 1--2 PeV energy) while track events,
most likely due to $\nu_\mu$ charge current $\nu N$ interactions,
constitute the rest. Among a total of 37 events about 15 could be due
to atmospheric neutrino ($6.6^{+5.9}_{-1.6}$) and muon ($8.4\pm 4.2$)
backgrounds.  A background-only origin of all 37 events has been
rejected at 5.7-$\sigma$ level \cite{Aartsen:2014gkd}. Therefore a
cosmic origin of a number of neutrino events is robust. The track
events have on average $\sim 1^\circ$ angular resolution, but the
dominant, shower events have much poorer angular resolution, $\sim
15^\circ$ on average \cite{Aartsen:2014gkd}, thus making them
unsuitable for astronomy.

Meanwhile the Pierre Auger Observatory (PAO)
\cite{PierreAuger:2014yba} and the Telescope Array (TA)
\cite{Abbasi:2014lda}, two of the world's largest operating cosmic-ray
detectors, have recently released UHECR data collected over more than
10-year and 5-year periods, respectively.  Together they have detected
16 events (6 by PAO \cite{PierreAuger:2014yba} and 10 by TA
\cite{Abbasi:2014lda}) with energies $\gtrsim 100$ EeV. The total
publicly available $\gtrsim 100$ EeV events including past experiments
is 33.  While lower-energy cosmic ray arrival directions are scrambled
by the Galactic and intergalactic magnetic fields, at $\gtrsim 60$ EeV
energies the arrival directions of UHECRs tend to be much better
correlated with their source directions and astronomy with charged
particles could be realized \cite{Sommers:2008ji}. Few degree angular
resolution can be achieved at these energies, which is much better
than the IceCube neutrino shower events and is comparable to the
neutrino track events.

The astrophysical sources of UHECRs with energy $\gtrsim 40$ EeV need
to be located within the so-called GZK volume \cite{Greisen:1966jv,
  Zatsepin:1966jv} in order to avoid serious attenuation of flux from
them due to interactions of UHECRs with photons from cosmic microwave
background (CMB) and extragalactic background light (EBL).  The
astrophysical sources of neutrinos, on the other hand, can be located
at large distances and still be detected provided their luminosity is
sufficiently high. However, because of weakly interacting nature of
neutrinos and limiting luminosity of astrophysical sources, only
nearby neutrino sources can be identified, thus making neutrino
astronomy possible.

We explore here a possibility that both UHECRs and IceCube cosmic
neutrino events are produced by the same astrophysical sources within
the GZK volume.  Since widely accepted Fermi acceleration mechanism of
cosmic rays at the sources take place over a large energy range, it is
natural that the same sources produce $\gtrsim 1$ PeV cosmic rays,
required to produce cosmic neutrinos observed with energies down to
$\sim 30$ TeV, and UHECRs with energy $\ge 40$ EeV.  We employ
invariant statistical method \cite{Virmani:2002xk, Razzaque:2001tp},
independent of coordinate systems, in order to study angular
correlation between cosmic neutrinos and UHECRs.  As far we know, this
is the first attempt to quantify such a correlation between the
IceCube neutrino and UHECR data sets. Existence of such a correlation
can provide clues to the origin of both cosmic neutrinos and UHECRs.
We search for astrophysical sources within the angular errors of
UHECRs which are correlated with the neutrino events in order to shed
lights on their plausible, common origins. Finally we calculate
required neutrino and cosmic-ray luminosities for the sources to
produce observed events, and compare these luminosities with their
observed X-ray and radio luminosities to check if they are viable
sources of both UHECRs and cosmic neutrinos.

The organization of this paper is the following. We describe neutrino
and UHECR data that we use in section \ref{data} and our statistical
method in section \ref{method}. The results of our correlation study
and source search along the directions of the correlated events are
given in section \ref{results}. Section \ref{results} also includes
calculation of neutrino and cosmic-ray luminosities of the correlated
sources from respective fluxes derived using data. We discuss our
results and implications of our findings in section \ref{discussion}.

\section{IceCube neutrino events and UHECR data}
\label{data}

\begin{figure}[tbp]
\centering 
\includegraphics[width=16cm]{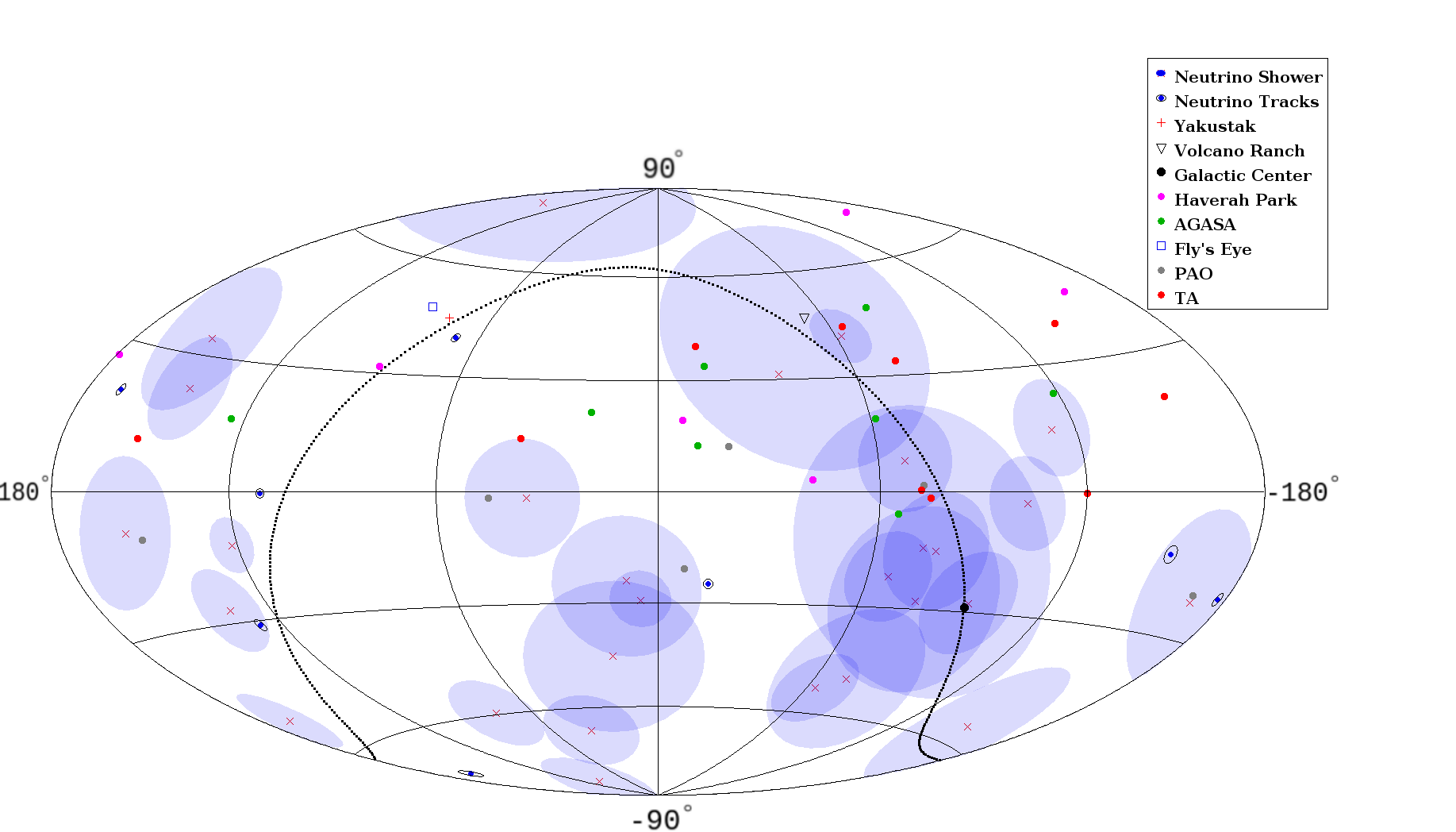}
\includegraphics[width=16cm]{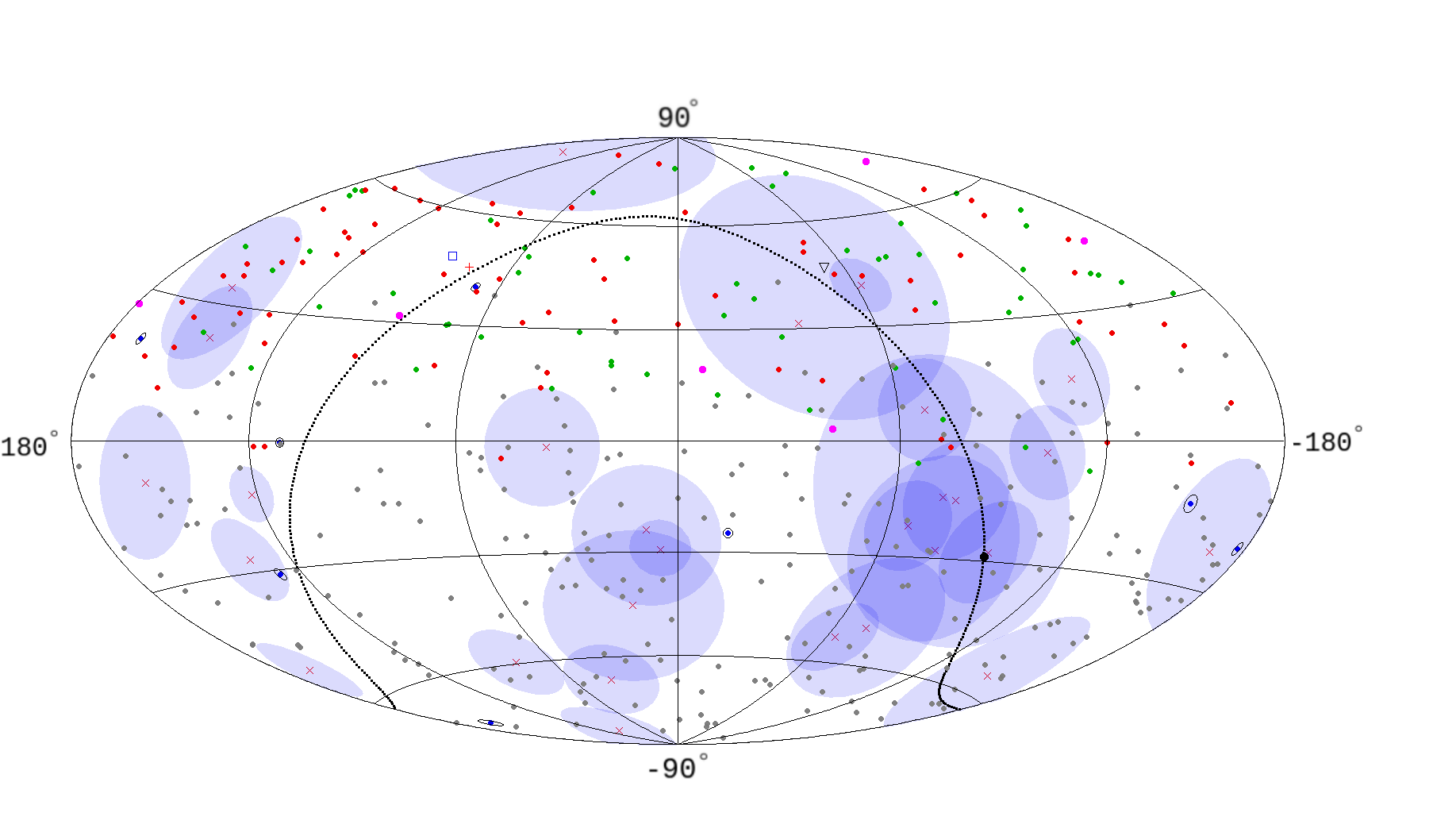}
\caption{\label{fig:skymap} Sky maps of the IceCube $>30$ TeV cosmic
  neutrino events with error circles and UHECR data in equatorial
  coordinates. The top panel shows UHECRs with energy $\ge 100$ EeV
  and the bottom panel shows all available data with energy $\ge 40$
  EeV. The black dotted line is the Galactic plane.}
\end{figure}

\begin{figure}[tbp]
\centering 
\includegraphics[width=16cm]{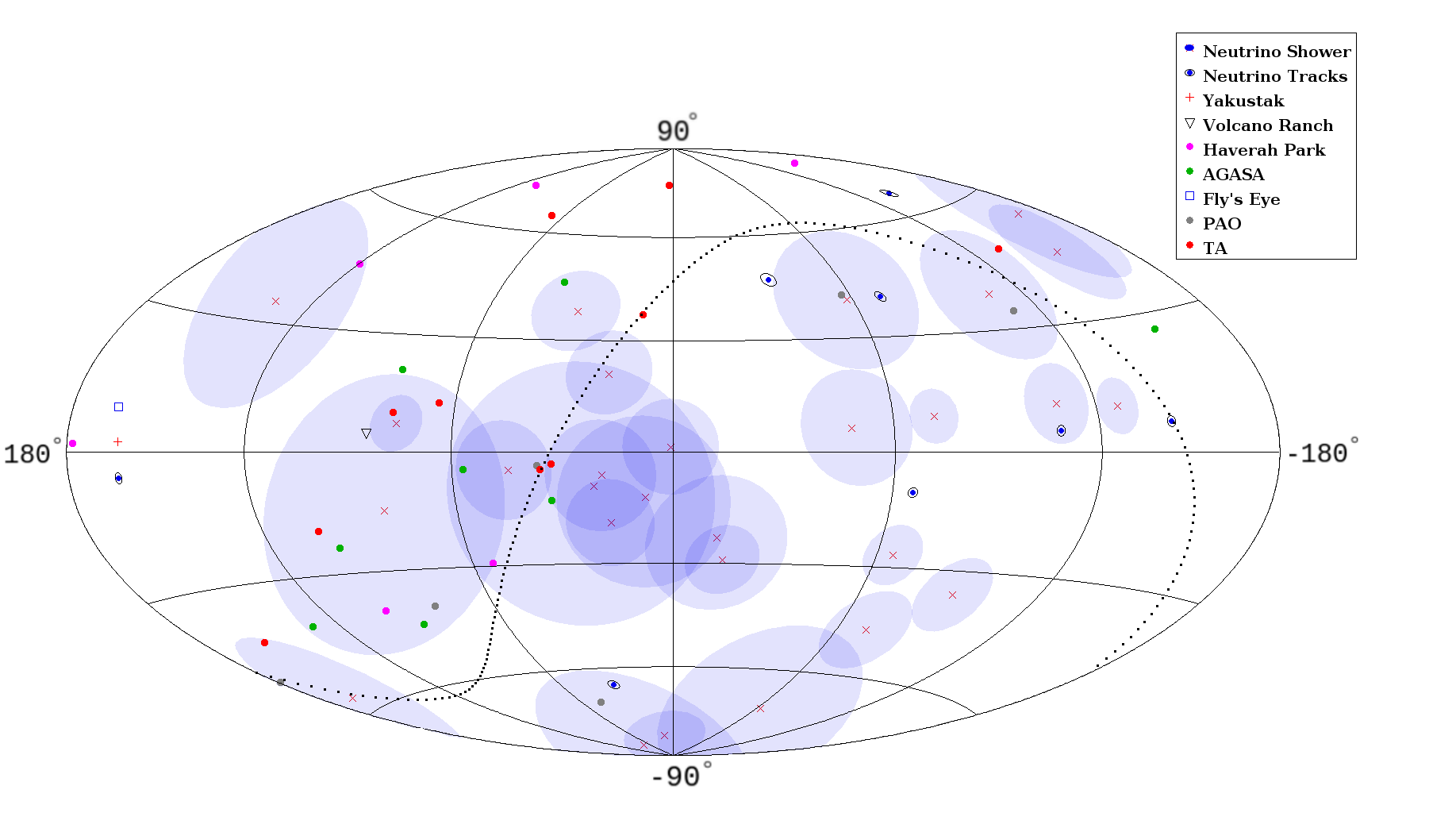}
\includegraphics[width=16cm]{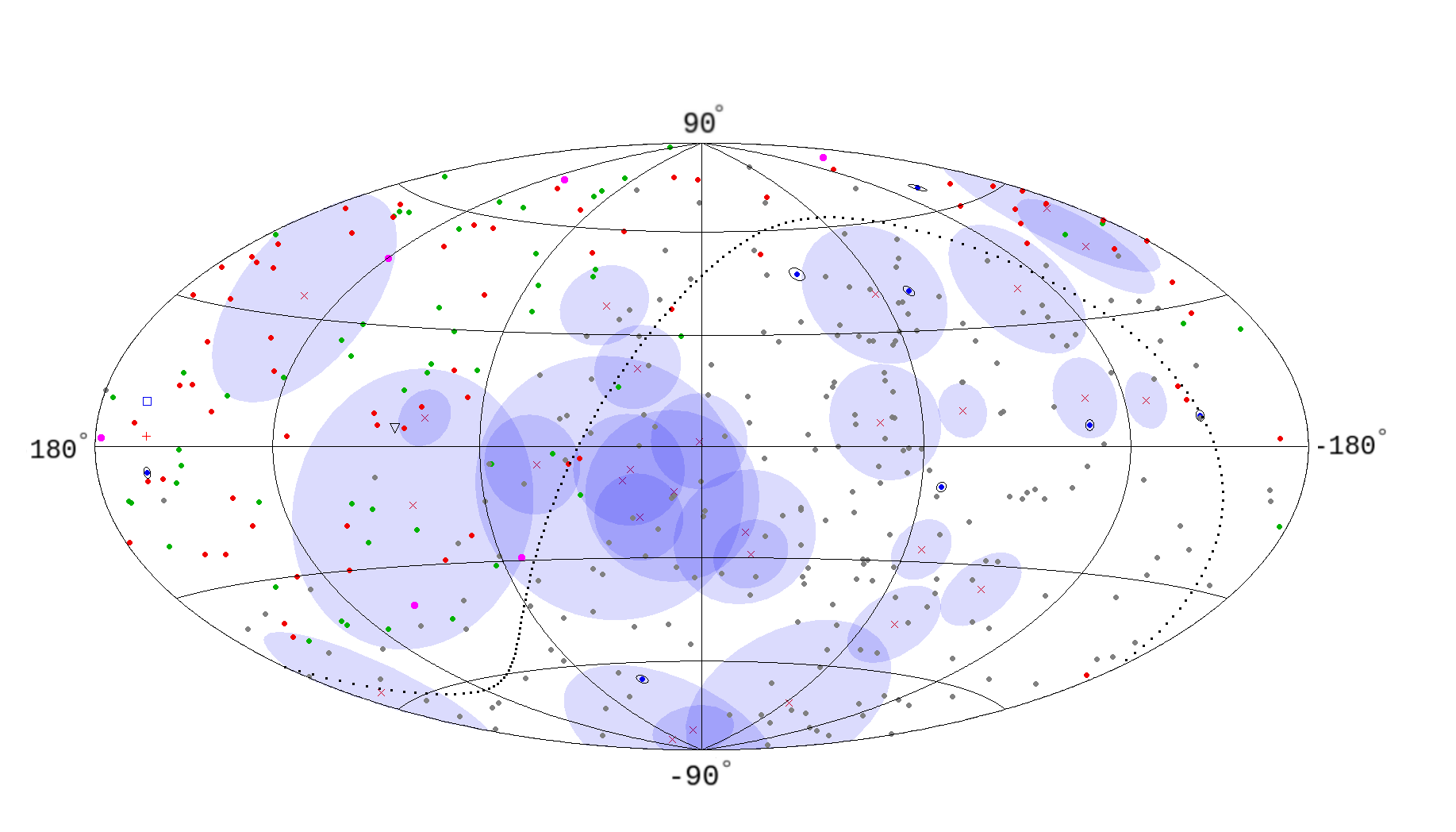}
\caption{\label{fig:skymap2} The same as Figure \ref{fig:skymap} but
  in Galactic coordinates.}
\end{figure}

We consider 35 IceCube neutrino events, collected over 988 days in the
$\sim 30$ TeV--2 PeV range, from ref.~\cite{Aartsen:2014gkd} to study
angular correlation with UHECRs. Two track events (event numbers 28
and 32) are coincident hits in the IceTop surface array and are almost
certainly a pair of atmospheric muon background events
\cite{Aartsen:2014gkd}. Therefore we excluded them from our analysis.
Figure \ref{fig:skymap} shows sky maps of 35 events in equatorial
coordinates with reported angular errors.  The majority (26) of the
events have arrival directions in the southern sky. Among the 9
northern hemisphere events, only 1 is at a declination $Dec >
41^\circ$ which happens to be a shower event \cite{Aartsen:2014gkd}.
The angular resolutions of the track events are $\lesssim 1.4^\circ$
and those of the shower events vary between $6.6^\circ$ and
$46.3^\circ$. Figure \ref{fig:skymap2} shows sky maps in Galactic
coordinates.

Figures \ref{fig:skymap} and \ref{fig:skymap2} also show sky maps of
available UHECR data with energies $\ge 100$ EeV (top panel) and $\ge
40$ EeV (bottom panel). The PAO and TA collaborations have published
data with energies above 52 EeV (231 events)
\cite{PierreAuger:2014yba} and 57 EeV (72 events)
\cite{Abbasi:2014lda}, respectively. Note that the PAO and TA are
located in the southern and northern hemisphere, respectively,
covering a declination range of $-90^\circ \le Dec \le 45^\circ$
\cite{PierreAuger:2014yba} and $-10^\circ \le Dec \le 90^\circ$
\cite{Abbasi:2014lda}. The angular resolutions for the PAO events with
energy $>10$ EeV is $<0.9^\circ$ \cite{Bonifazi:2009ma} while for the
TA events with energy $> 57$ EeV it is between $1.0^\circ$ and
$1.7^\circ$ \cite{Abbasi:2014lda}. Note that the $\ge 40$ EeV data
sample is incomplete.  Only the AGASA experiment has published data
above 40 EeV (40 events) covering a declination range $-10^\circ \le
Dec \le 90^\circ$ and with angular resolution $< 2^\circ$
\cite{Takeda:1999sg}. Only $\ge 100$ EeV data are available from the
other past experiments: Haverah Park \cite{Nagano:2000ve, Ave:2000nd},
Yakutsk \cite{Nagano:2000ve}, Volcano Ranch \cite{Nagano:2000ve} and
Fly's eye \cite{Nagano:2000ve}. Note that these were all northern
hemisphere experiments. We could not include 13 events with energy $>
56$ EeV from the HiRes experiment as the energies of the individual
events are not available \cite{Abbasi:2008md}.

We list in Table \ref{tab:i} all available UHECR data with energy $\ge
100$ EeV. This includes 6 events from Haverah Park, 1 event from
Yakutsk, 8 events from AGASA, 1 event from Volcano Ranch, the
highest-energy (320 EeV) event from Fly's eye, 6 events from PAO and
10 events from TA. We use this list and sublists with PAO and TA data
separately to study correlation with cosmic neutrino events.
In addition to $\ge 100$ EeV data, we also explore
  energy-dependence of correlation by choosing different energy cuts,
  $\ge 80$ EeV and $\ge 60$ EeV, in UHECR data from the PAO, TA and
  AGASA . Above 80 EeV (60 EeV) there are 22 (136) UHECR events from
  PAO, 20 (60) from TA and 11 (22) from AGASA. We note that energy
  calibration across the experiments can vary by as much as $\sim
  30\%$ (see, e.g., ref.~\cite{Bluemer:2009zf}). We discuss this issue
  in Sec.~\ref{calibration}.

\begin{table}[tbp]
\centering
\begin{tabular}{|l|c|c|c|c|}
\hline
Experiment & Reference & Energy (EeV) & RA ($^\circ$) & Dec ($^\circ$) \\
\hline
Haverah Park	& \cite{Nagano:2000ve}  & 101 & 201 & 71 \\
Haverah Park	& \cite{Nagano:2000ve}  & 116 & 353 & 19 \\
Haverah Park	& \cite{Nagano:2000ve}  & 126 & 179 & 27 \\
Haverah Park	& \cite{Nagano:2000ve}  & 159 & 199 & 44 \\
Haverah Park	& \cite{Ave:2000nd} & 123 & 318.3 & 3.0 \\
Haverah Park	& \cite{Ave:2000nd} & 115 & 86.7 & 31.7 \\
Yakutsk		& \cite{Nagano:2000ve}  & 110 & 75.2 & 45.5 \\
AGASA 		& \cite{Takeda:1999sg} & 101 & 124.25 & 16.8 \\
AGASA		& \cite{Takeda:1999sg} & 213 & 18.75 & 21.1 \\
AGASA		& \cite{Takeda:1999sg} & 106 & 281.25 & 48.3 \\
AGASA		& \cite{Takeda:1999sg} & 144 & 241.5 & 23.0 \\
AGASA		& \cite{Takeda:1999sg} & 105 & 298.5 & 18.7 \\
AGASA		& \cite{Takeda:1999sg} & 150 & 294.5 & $-5.8$ \\
AGASA		& \cite{Takeda:1999sg} & 120 & 349.0 & 12.3 \\
AGASA		& \cite{Takeda:1999sg} & 104 & 345.75 & 33.9 \\
Volcano Ranch	& \cite{Nagano:2000ve}  & 135 & 306.7 & 46.8 \\
Fly's eye	& \cite{Nagano:2000ve}  & 320 & $85.2\pm 0.5$ & $48.0^{+5.2}_{-6.3}$ \\
Pierre Auger	& \cite{PierreAuger:2014yba} & 108.2 & 45.6 & $-1.7$ \\
Pierre Auger	& \cite{PierreAuger:2014yba} & 127.1 & 192.8 & $-21.2$ \\
Pierre Auger	& \cite{PierreAuger:2014yba} & 111.8 & 352.6 & $-20.8$ \\
Pierre Auger	& \cite{PierreAuger:2014yba} & 118.3 & 287.7 & $1.5$ \\
Pierre Auger	& \cite{PierreAuger:2014yba} & 100.1 & 150.1 & $-10.3$ \\
Pierre Auger	& \cite{PierreAuger:2014yba} & 118.3 & 340.6 & $12.0$ \\
Telescope Array & \cite{Abbasi:2014lda} & 101.4 & 285.74 & $-1.69$ \\
Telescope Array & \cite{Abbasi:2014lda} & 120.3 & 285.46 & 33.62 \\
Telescope Array & \cite{Abbasi:2014lda} & 139.0 & 152.27 & 11.10 \\
Telescope Array & \cite{Abbasi:2014lda} & 122.2 & 347.73 & 39.46 \\
Telescope Array & \cite{Abbasi:2014lda} & 154.3 & 239.85 & $-0.41$ \\
Telescope Array & \cite{Abbasi:2014lda} & 162.2 & 205.08 & 20.05 \\
Telescope Array & \cite{Abbasi:2014lda} & 124.8 & 295.61 & 43.53 \\
Telescope Array & \cite{Abbasi:2014lda} & 135.5 & 288.30 & 0.34 \\
Telescope Array & \cite{Abbasi:2014lda} & 101.0 & 219.66 & 38.46 \\
Telescope Array & \cite{Abbasi:2014lda} & 106.8 & 37.59 & 13.89 \\
\hline
\end{tabular}
\caption{\label{tab:i} Available UHECR data with energy $\gtrsim 100$ EeV from various experiments.}
\end{table}

\section{Statistical method and data analyses}
\label{method}

To study correlation between cosmic neutrinos and UHECRs, we map the
Right Ascension and Declination $(RA, Dec)$ of the event directions
into unit vectors on a sphere as
$$
{\hat x} = (\sin\theta \cos\phi, \sin\theta \sin\phi, \cos\theta)^T,
$$
where $\phi = RA$ and $\theta = \pi/2 - Dec$. Scalar product of the
neutrino and UHECR vectors $({\hat x}_{\rm neutrino}\cdot {\hat
  x}_{\rm UHECR})$ therefore is independent of the coordinate system.
The angle between the two vectors
\begin{equation}
\label{gamma}
\gamma = \cos^{-1} ({\hat x}_{\rm neutrino}\cdot 
{\hat x}_{\rm UHECR}),
\end{equation}
is an invariant measure of the angular correlation between the
neutrino and UHECR arrival directions \cite{Virmani:2002xk,
  Razzaque:2001tp}. Following ref.~\cite{Virmani:2002xk} we use a
statistic made from invariant $\gamma$ for each neutrino direction
${\hat x}_i$ and UHECR direction ${\hat x}_j$ pair as
\begin{equation}
\label{delta}
\delta\chi^2_i = {\rm min}_j (\gamma_{ij}^2/\delta\gamma_i^2),
\end{equation}
which is minimized for all $j$. Here $\delta\gamma_i$ is the
1-$\sigma$ angular resolution of the neutrino events. We use the exact
resolutions reported by the IceCube collaboration for each event
\cite{Aartsen:2014gkd}.

A value $\delta \chi^2_i \le 1$ is considered a ``good match'' between
the $i$-th neutrino and an UHECR arrival directions. We exploit
distributions of all $\delta\chi^2_i$ statistic to study angular
correlation between IceCube neutrino events and UHECR data.  The
distribution with observed data giving a number of ``hits'' or $N_{\rm
  hits}$ with $\delta\chi^2 \le 1$ therefore forms a basis to claim
correlation. Note that in case more than one UHECR directions are
within the error circle of a neutrino event, the $\delta\chi^2$ value
for UHECR closest to the neutrino direction is chosen in this method.

We estimate the significance of any correlation in data by comparing
$N_{\rm hits}$ with corresponding number from null distributions.
We construct two null distributions, in one case we
  randomize only the $RA$ of UHECRs, keeping their $Dec$ the same as
  in data; and in the second case we also randomize $Dec$ according to
  the zenith-angle depended sky exposure of the UHECR experiments
  \cite{Sommers:2000us}, affecting the declination distributions of
  UHECR data. We call these two null distributions as the {\it
    semi-isotropic null} and {\it exposure-corrected null},
  respectively. The {\it semi-isotropic null} is a quick-way to check
  significance while the {\it exposure-corrected null} is accurate
  when information on particular experiments are available. In both
  cases we perform 100,000 realizations of drawing random numbers to
  assign new $RA$ and $Dec$ values for each event to construct
  $\delta\chi^2$ distributions in the same way as done with real data.
  We find that the two null distributions are in good agreement with
  each other in most cases.

We calculate statistical significance of correlation
  in real data or $p$-value (chance probability) using frequentists'
  approach. We count the number of times we get a random data set that
  gives equal or more hits than the $N_{\rm hits}$ in real data within
  $\delta\chi^2 \le 1$ bin.  Dividing this number with the total
  number of random data sets generated (100,000) gives us the
  $p$-value. We cross-check this $p$-value by calculating the Poisson
  probability of obtaining $N_{\rm hits}$ within $\delta\chi^2 \le 1$
  bin given the corresponding average hits expected from the null
  distribution.  These two chance probabilities are in good
  agreement.

\section{Results}
\label{results}

\subsection{Correlations between neutrinos and UHECRs}

\begin{figure}[tbp]
\centering 
\includegraphics[width=.9\textwidth, trim=0 0 0 0,clip]{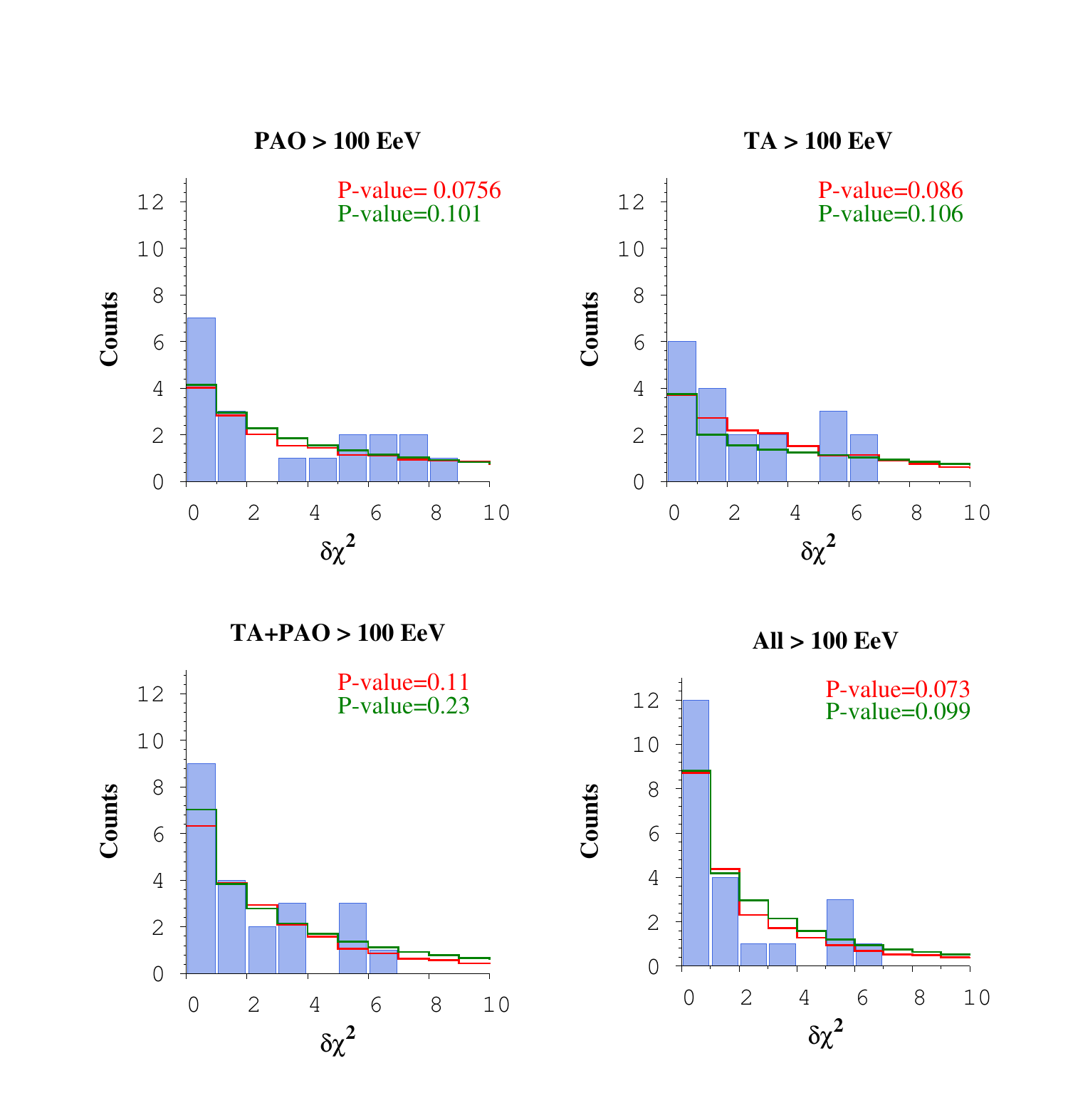}
\caption{\label{fig:100EeV} Distributions of $\delta\chi^2$ found in
  observed data (blue, filled histograms) and in simulated data
  corresponding to the {\it semi-isotropic null} (red, open
  histograms) and the {\it exposure-corrected null} (green, open
  histograms). The histograms have been truncated at $\delta\chi^2 =
  10$ for better display. Significances ($p$-values) have been
  calculated for the $0 \le \delta\chi^2 \le 1$ bins. }
\end{figure}

We apply our statistical method separately to the PAO and TA data,
to a combination of the both and to all available UHECR data above 100
EeV from all experiments (see Table \ref{tab:i}).  The results are
shown in the histograms of Figure \ref{fig:100EeV}.  The counts in the
$0 \le \delta\chi^2 \le 1$ bins for the blue, filled histograms
correspond to the number of correlated neutrino events with UHECRs.
Counts in other bins are due to distant pairs of the neutrino events
and UHECRs and are uninteresting for us.  The counts for the red
(green), open histogram in the same bins correspond to the expected
number of correlated neutrino events from the {\it semi-isotropic
  null} ({\it exposure-corrected null}), after averaging over 100,000
simulated data sets with random UHECR positions. Both null
distributions give similar results.


Figure \ref{fig:100EeV} also shows $p$-values or the probability of
finding the correlated events ($0 \le \delta\chi^2 \le 1$) in observed
data as a fluctuation of the randomly distributed UHECRs in the sky.
The probability $1-p \approx 90\%$ is the confidence level (CL) that
the IceCube neutrino events and all available UHECR data with energy
$\ge 100$ EeV are correlated. A correlation with similar CL exists
between the neutrino and { PAO-only data sets and between
  the neutrino and TA-only data sets}. The Poisson probability of
obtaining $N_{\rm hits} = 7$ in PAO data when 4 are expected from the
{{\it semi-isotropic null distribution} and $N_{\rm
    hits} = 6$ in TA data with 3.8 expected from the same null
  distribution are $\approx 0.06$}, in very good agreement with our
$p$-values.  Similarly, for the combined data set of all UHECRs $>
100$ EeV, $N_{\rm hits} = 12$ when expected value is 8.8 corresponds
to a Poisson probability of $\approx 0.07$, again in very good
agreement with our $p$-value.

We remind the readers that the counts in the $\delta\chi^2$
distributions with TA+PAO data is not the algebraic sum of the counts
in the distributions with TA and PAO data separately. This is because
our $\delta\chi^2$ statistic choose the nearest UHECR data point even
if more than one are present within the error circle of a neutrino
event. The same is true for distribution with all UHECRs. We list the
correlated events in Table \ref{tab:ii} against the neutrino event
numbers in ref.~\cite{Aartsen:2014gkd}. Note that we list all UHECRs
giving $\delta\chi^2 \le 1$ in the table. There are 7 UHECRs which are
correlated with 2 or more neutrino events. None of the correlated
neutrinos are PeV neutrino events.

\begin{figure}[tbp]
\centering 
\includegraphics[width=.9\textwidth, trim=0 0 0 0,clip]{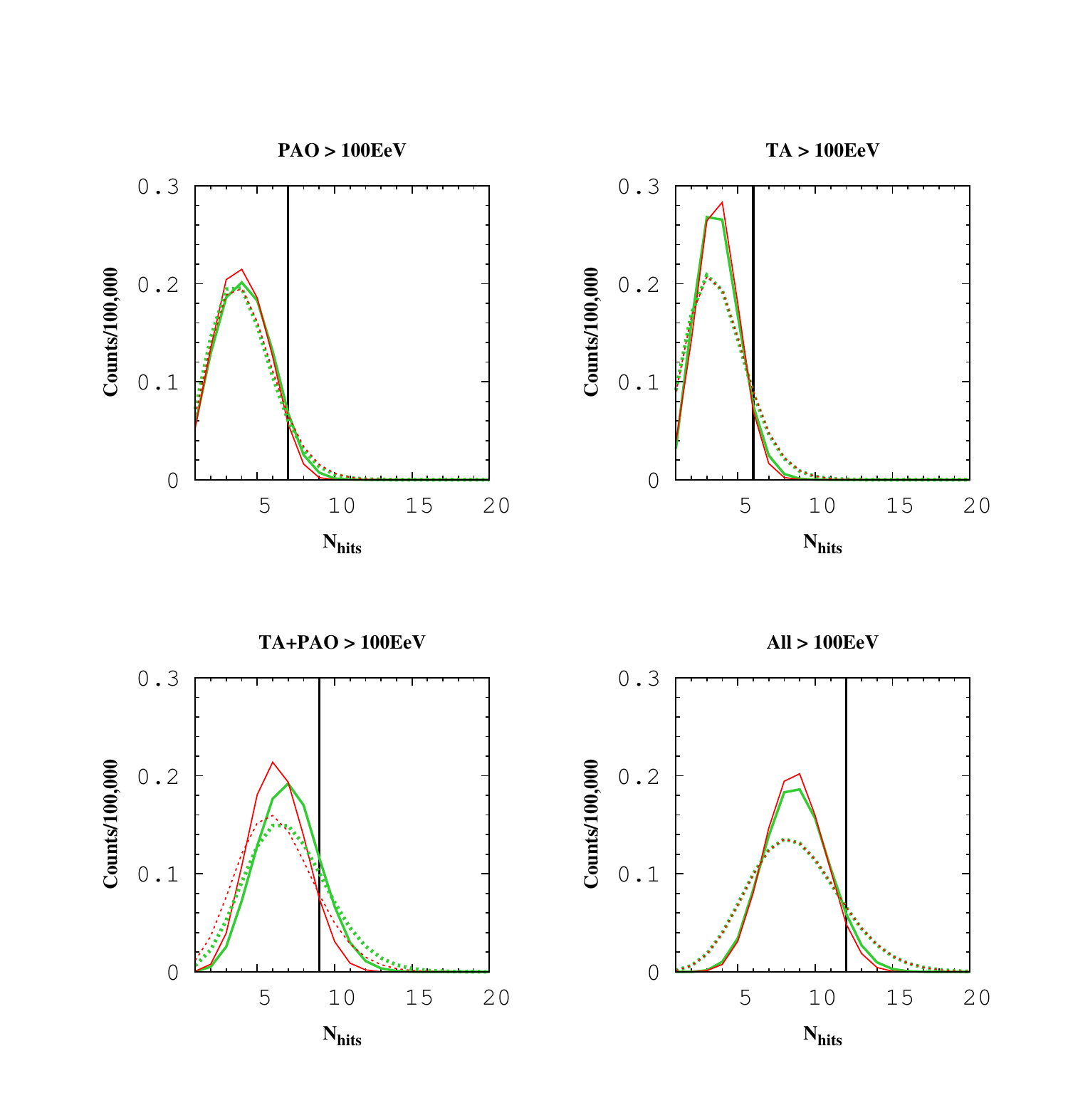}
\caption{\label{fig:distribution} Comparisons between the $N_{\rm
    hits}$ distributions in the $\delta\chi^2 \le 1$ bins of
  Fig.~\ref{fig:100EeV} obtained from the {\it semi-isotropic} (red
  solid lines) and {\it exposure-corrected} (green solid lines) null
  distributions. Also plotted are the Poisson distributions (dotted
  lines) for the average values of the respective null distributions
  in the $\delta\chi^2 \le 1$ bins of Fig.~\ref{fig:100EeV}. The
  vertical lines are the observed $N_{\rm hits}$ values in data.}
\end{figure}

{\ Figure \ref{fig:distribution} shows a comparison between
  the two null distributions for UHECRs with energy $\ge 100$ EeV in
  Fig.~\ref{fig:100EeV} by using the $N_{\rm hits}$ within the $\delta
  {\chi}^2 \le 1$ bin from simulations for both the null
  distributions. The two null distributions agree well in all cases
  except for the combined analysis of the TA and PAO data.  A
  comparison with Poisson distribution with frequency for the
  corresponding cases are also shown. For PAO UHECRs >100 EeV the two
  null distributions follow the respective Poisson distributions but
  not for the other cases. The black vertical line represents the
  observed $N_{\rm hits}$.}

\begin{table}[tbp]
\centering
\begin{tabular}{|l|c|c|c|c|c|}
\hline
$\nu$ event no.~\cite{Aartsen:2014gkd} & $\delta\chi^2$ & Energy (EeV) 
& RA ($^\circ$) & Dec ($^\circ$) & Experiment \\
\hline
1  & 0.41 & 108.2  & 45.6   & $-1.7$  & PAO \\
   & 0.95 & 106.8  & 37.59  & 13.9    & TA \\
2  & 0.97 & 150    & 294.5  & $-5.8$  & AGASA \\
11 & 0.10 & 100.1  & 150.1  & $-10.3$ & PAO \\
16 & 0.006 & 127.1  & 192.8  & $-21.2$ & PAO \\
17 & 0.77 & 144    & 241.5  & 23.0    & AGASA \\
21 & 0.55 & 111.8  & 352.6  & $-20.8$ & PAO \\
24 & 0.78 & 101.4  & 285.74 & $-1.7$  & TA \\
   & 0.97 & 150    & 294.5  & $-5.8$  & AGASA \\   
25 & 0.06 & 150    & 294.5  & $-5.8$  & AGASA \\
   & 0.07 & 101.4  & 285.74 & $-1.7$  & TA \\
   & 0.10 & 135.5  & 288.3  & 0.34    & TA \\
   & 0.12 & 118.3  & 287.7  & 1.5     & PAO \\    
   & 0.58 & 105    & 298.5  & 18.7    & AGASA\\         
   & 0.62 & 123    & 318.3  & 3       & Haverah Park \\      
29 & 0.18 & 124.8  & 295.6  & 43.52   & TA \\
31 & 0.35 & 101    & 201    & 71      & Haverah Park \\
33 & 0.34 & 118.3  & 287.7  & 1.5     & PAO \\
   & 0.40 & 135.5  & 288.3  & 0.34    & TA \\
   & 0.74 & 101.4  & 285.74 & $-1.7$  & TA \\        
   & 0.84 & 105    & 298.5  & 18.7    & AGASA \\     
34 & 0.20 & 104    & 345.75 & 34      & AGASA \\
   & 0.22 & 135    & 306.7  & 46.8    & Volcano Ranch \\
   & 0.25 & 122.2  & 347.7  & 39.46   & TA \\
   & 0.34 & 118    & 340.6  & 12      & PAO \\
   & 0.34 & 124.8  & 295.61 & 43.53   & TA \\
   & 0.36 & 105    & 298.5  & 18.7    & AGASA \\
   & 0.45 & 123    & 318.3  & 3       & Haverah Park \\
   & 0.47 & 116    & 353    & 19      & Haverah Park \\
   & 0.50 & 120    & 349    & 12.3    & AGASA \\
   & 0.55 & 120.3  & 285.5  & 33.62   & TA \\   
   & 0.71 & 134    & 281.25 & 48.3    & AGASA \\
\hline
\end{tabular}
\caption{\label{tab:ii} 
  IceCube cosmic neutrino events correlated with UHECRs above 100 EeV.}
\end{table}

\begin{figure}[tbp]
\centering 
\includegraphics[width=.9\textwidth, trim=0 0 0 0,clip]{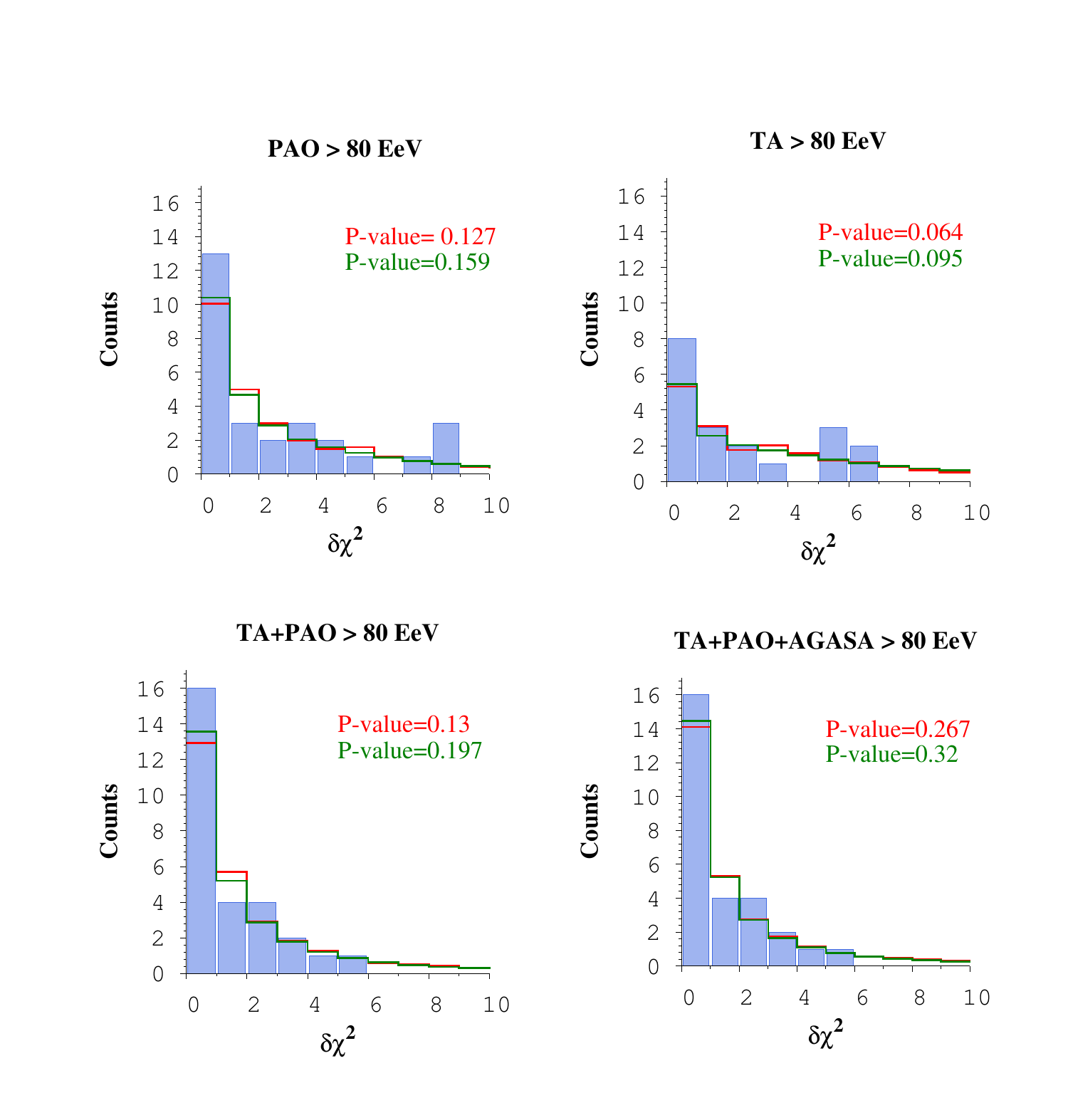}
\caption{\label{fig:80EeV} The same as in Figure \ref{fig:100EeV} but
  for UHECRs with energy $\ge 80$ EeV.}
\end{figure}

We do the same statistical analysis with UHECRs above 80 EeV.  The
results are shown in Figure \ref{fig:80EeV}.  There are 6 new
correlations in the PAO-only data. Only 10 is expected from our null
distributions as compared to 13 total in data. This reduces
correlation between the IceCube and PAO data to $\approx 84\% -88\%$
CL.  A list of UHECRs from the PAO correlated with the neutrino events
is given in Table \ref{tab:iii}. Note that the 2 PeV neutrino event
(event number 35), the highest observed energy so far
\cite{Aartsen:2014gkd}, is now correlated with a 89.1 EeV PAO event.
The other two neutrino events with energy $\approx 1$ PeV each still
remain uncorrelated with UHECRs with energy $\ge 80$ EeV. There are 6
UHECRs in Table \ref{tab:iii} which are correlated with more than one
neutrino events. In particular an 82 EeV PAO event is correlated with
4 neutrino events.

\begin{table}[tbp]
\centering
\begin{tabular}{|l|c|c|c|c|}
\hline
$\nu$ event no.~\cite{Aartsen:2014gkd} & $\delta\chi^2$ & Energy (EeV) 
& RA ($^\circ$) & Dec ($^\circ$)\\
\hline
1  & 0.41  & 108.2  & 45.6   & $-1.7$ \\
2  & 0.002 & 80.9   & 283.7  & $-28.6$ \\
7  & 0.85  & 83.8   & 26.8   & $-24.8$ \\
11 & 0.10  & 100.1  & 150.1  & $-10.3$ \\
15 & 0.6   & 82.3   & 287.7  & $-64.9$ \\
16 & 0.006 & 127.1  & 192.8  & $-21.2$ \\
   & 0.52  & 84.7   & 199.7  & $-34.9$ \\
21 & 0.55  & 111.8  & 352.6  & $-20.8$ \\
21 & 0.6   & 83.8   & 26.8   & $-24.8$ \\
22 & 0.85  & 80.9   & 283.7  & $-28.6$ \\
24 & 0.77  & 80.9   & 283.7  & $-28.6$ \\
25 & 0.095 & 80.2   & 283.7  & $-28.6$ \\
   & 0.12  & 118.3  & 287.7  & 1.5 \\
   & 0.61  & 82     & 299    & $19.4$ \\
   & 0.62  & 80.2   & 271.1  & $19.0$ \\
   & 0.67  & 81.4   & 308.8  & 16.1 \\
33 & 0.34  & 82     & 287.7  & 1.5 \\ 
   & 0.95  & 82     & 299    & 19.4 \\
34 & 0.22  & 81.4   & 308.8  & 16.1 \\
   & 0.34  & 82     & 299.   & 19.4 \\
   & 0.34  & 118.3  & 340.6  & 12 \\
   & 0.6   & 89     & 349.9  & 9.3 \\
35 & 0.96  & 89.1   & 218.8  & $-70.8$ \\
\hline
\end{tabular}
\caption{\label{tab:iii} IceCube cosmic neutrino events correlated with UHECRs detected by PAO above 80 EeV.}
\end{table}


{\ Lowering the UHECR energy lower limit to 80 EeV adds 2
  new correlated events in the case of TA-only data with a total of 8
  as compared to $\approx 5.6$ expected from both the null
  distributions, giving a CL of $\approx 90\%$.  Combining the PAO and
  TA data results in similar significance as obtained from individual
  data sets.  Combining the PAO, TA and AGASA data reduces the
  significance of correlation.}

\begin{figure}[tbp]
\centering 
\includegraphics[width=.9\textwidth, trim=0 0 0 0,clip]{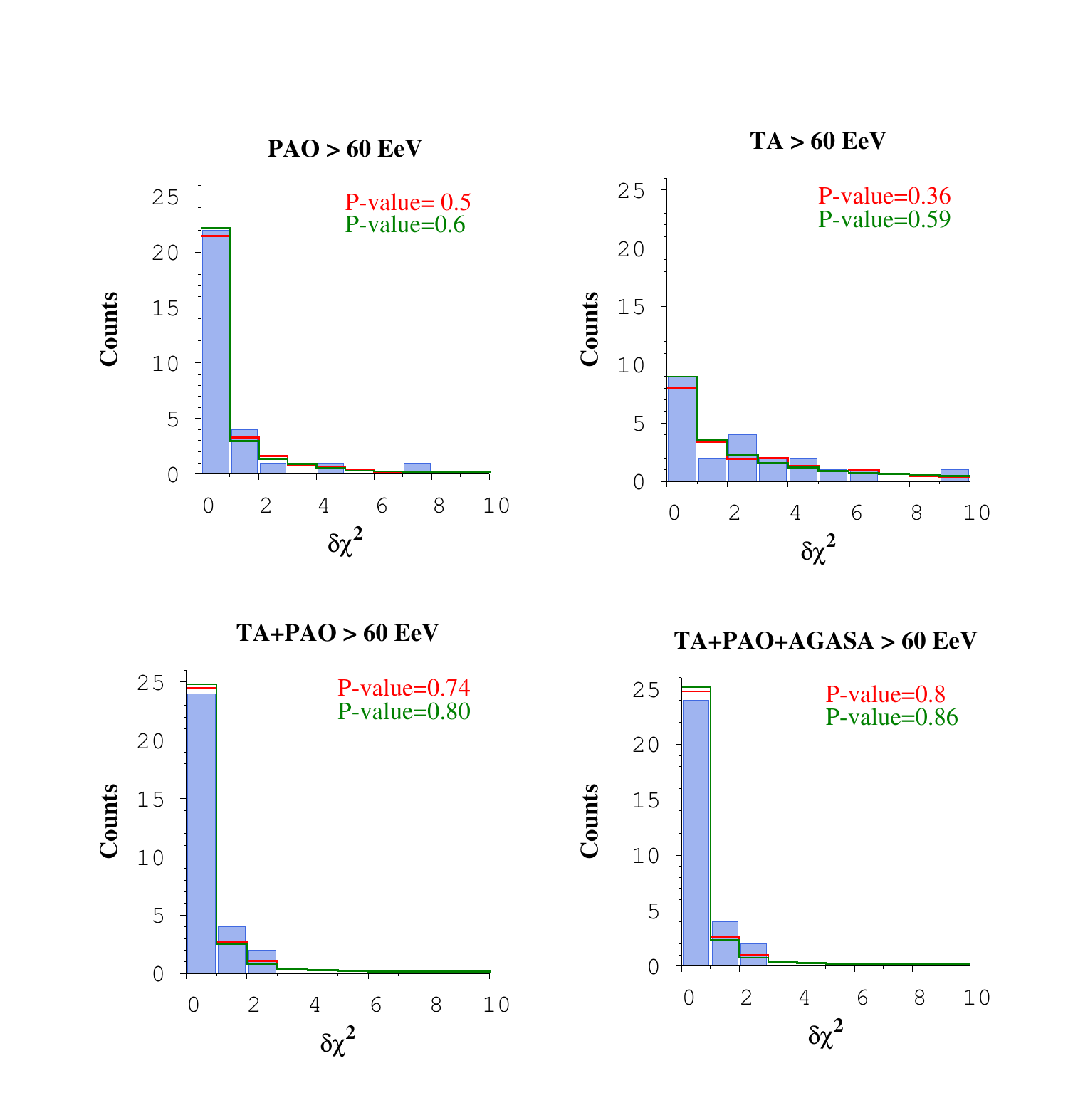}
\caption{\label{fig:60EeV} The same as in Figure \ref{fig:100EeV} but
  for UHECRs with energy $\ge 60$ EeV.  }
\end{figure}

Further lowering the UHECR energy lower limit to 60 EeV gives no
significant correlation between the IceCube cosmic neutrino data and
UHECR data. Figure \ref{fig:60EeV} shows that the number of correlated
events in data is very similar to those expected from the null
distributions in all cases. Such a loss of significance is expected
when there is no real correlation between the data sets.

\subsection{Energy calibrated events}
\label{calibration}

As we noted earlier, energy calibration among different UHECR
experiments is a widely-debated issue. If UHECR flux is uniform over
the whole sky and each experiment measures the same primary particles'
energy then the number of UHECR events should be proportional to the
exposures of the experiments. Therefore it is difficult, in
particular, to reconcile the 10 TA events at $> 100$ EeV compared to
the 6 from PAO, which has 20 times more exposure than TA. There are
many energy rescaling procedure suggested among experiments (see,
e.g.,~refs.~\cite{Aab:2014ila,Dawson:2013wsa,Ivanov:2010vp}) to bring
their respective measured fluxes close to each other, mostly at the
``ankle'' regime. Even these procedures cannot reconcile number of
events, after exposure corrections, above 100 EeV among different
experiments. It is plausible that the energy rescaling factors
themselves are energy dependent, differing from the ankle regime to
the GZK regime.  Reconstructing such energy-dependent rescaling
factors is beyond the scope of this paper. We hope the experimental
collaborations will provide such factors in future.

\begin{figure}[tbp]
\centering 
\includegraphics[width=.7\textwidth, trim=0 0 0 0,clip]{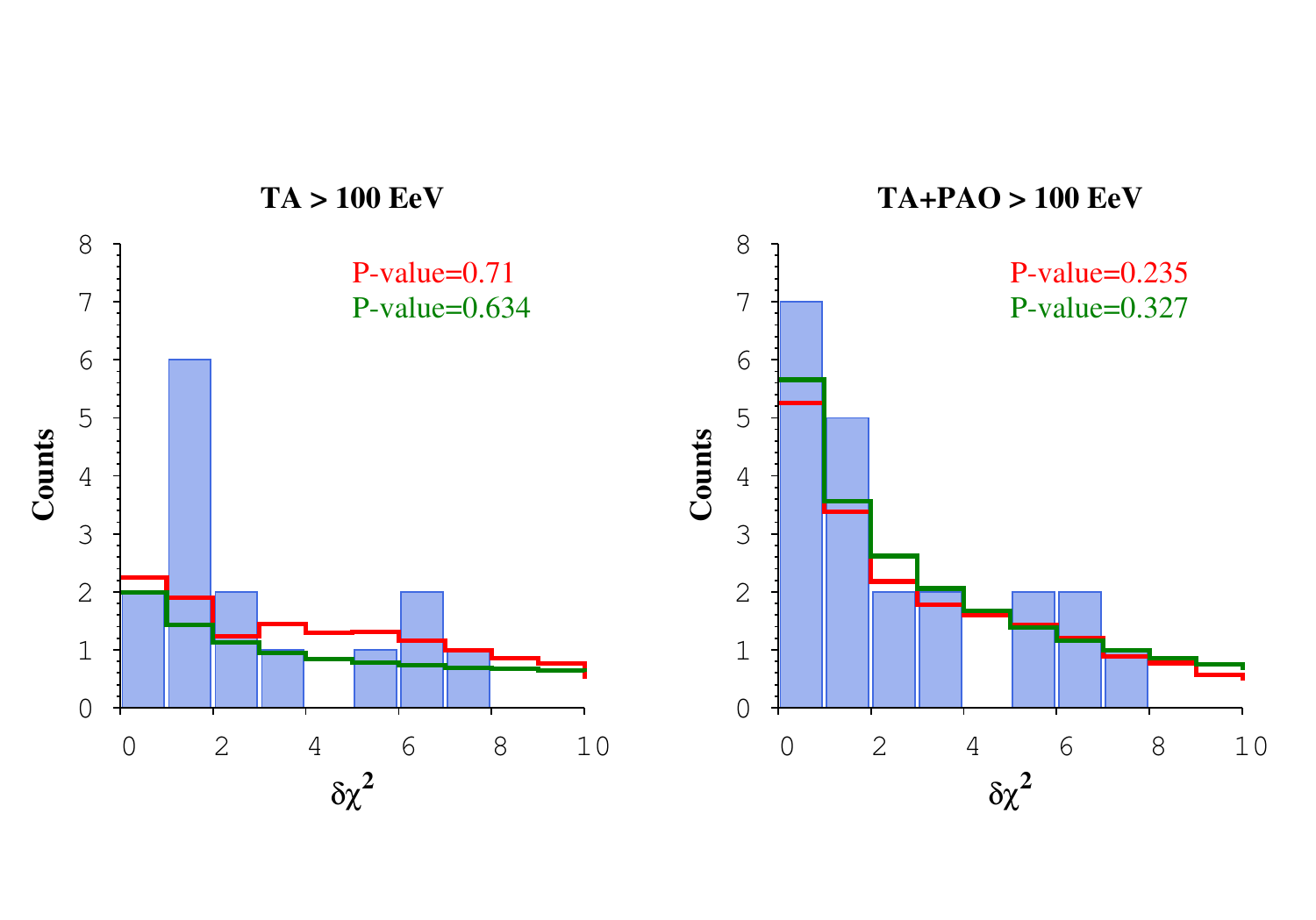}
\caption{\label{fig:TAcalibrated} The same as in Figure \ref{fig:100EeV}
  upper-right and bottom-left panels but the energies for the TA
  events have been reduced by $25\%$ compared to the PAO events.  }
\end{figure}

To illustrate the energy rescaling effect on our correlation study, we
adopt a recent procedure in ref.~\cite{Aab:2014ila} which is based on
a joint PAO and TA analysis of UHECRs from an overlapping region in
the sky.  We decrease the energies of the TA events by 25$\%$ but keep
the energies of the PAO events unchanged \cite{Aab:2014ila}.  So the
number of UHECRs events above 100 EeV from TA is now 4, out of which
only {\ 2} correlates with the neutrino events in the
$\delta\chi^2\le 1$ bin (see Fig.~\ref{fig:TAcalibrated}).
Interestingly, however, there are now 6 counts in the $1\le
\delta\chi^2\le 2$ bin which corresponds to a Poisson probability of
{\ 0.0108 according to the {\it semi-isotropic null}}. A
combined analysis of the TA and PAO data considering the above energy
rescaling, gives no significant correlation with neutrino data.

\subsection{Astrophysical source search}

We search for astrophysical source candidates for UHECRs which are
correlated with IceCube cosmic neutrino events, assuming both are
produced by the same sources. We use data from Tables \ref{tab:ii} and
\ref{tab:iii} for this purpose. The experimental angular resolution of
the UHECRs is of the order of $1^\circ$. However, Galactic and
intergalactic magnetic field can deflect them by more than a few
degrees from their source directions. The deflection angle in the
intergalactic random magnetic field \cite{Waxman:1996zn} with strength
$B_{\rm rdm}$ and coherence length $\lambda_{\rm coh}$ is
\begin{equation}
\delta\theta_{\rm IG} \approx 1.1^\circ Z
\left( \frac{E_{\rm cr}}{100~{\rm EeV}}\right)^{-1}
\left( \frac{B_{\rm rdm}}{1~{\rm nG}}\right)
\left( \frac{D}{200~{\rm Mpc}}\right)^{1/2}
\left( \frac{\lambda_{\rm coh}}{100~{\rm kpc}}\right)^{1/2}
\label{IGdeflection}
\end{equation}
where $Z$ and $E_{\rm cr}$ are the charge and energy of the UHECR and
$D$ is the distance to the source. The deflection angle in a
small-scale Galactic random magnetic field, using
Eq.~(\ref{IGdeflection}) with $B_{\rm rdm} = 1\mu$G, $\lambda_{\rm
  coh} = 100$~pc is much smaller, $\delta\theta_{\rm G} \approx
0.2^\circ Z$, for $E_{\rm cr}=100$~EeV and $D=10$~kpc. However, the
deflection angle in the large-scale regular component of the Galactic
magnetic field in the disk and in the halo can be larger, $\sim
1^\circ$--$3^\circ$ \cite{Farrar:2012gm, He:2014mqa}. Hereafter we
assume that UHECRs with energy $\ge 80$ EeV are dominantly protons
\footnote{Note that the mass composition measurement by the PAO
  collaboration by using shower maxima, which favors heavy nuclei as
  primaries, is done up to an energy $\sim 60$ EeV only
  \cite{ThePierreAuger:2013eja}.}. We also chose a conservative source
search region of $3^\circ$ around the directions of UHECRs which are
correlated with cosmic neutrino events.

We also limit our source search within a comoving volume with its
radius set by the GZK effect of $p_{\rm UHECR} + \gamma_{\rm CMB}$
interactions and corresponding energy losses by UHECR protons
\cite{Greisen:1966jv, Zatsepin:1966jv}. A crude estimate of the
mean-free-path for this interaction can be obtained from the number
density of CMB photons with $2.73$ K temperature in the local
universe, which is
\begin{equation}
\epsilon n(\epsilon) = \frac{1.32\times 10^4 (\epsilon/{\rm meV})^3}
{\exp\left[ 4.25(\epsilon/{\rm meV}) \right] - 1} ~{\rm cm}^{-3}.
\label{cmb_density}
\end{equation}
Thus the number density is 244~cm$^{-3}$ at the peak photon energy
$\epsilon = 2.82 k_B(2.73~{\rm K}) = 0.66$~meV, where $k_B =
8.62\times 10^{-5}$~eV~K$^{-1}$ is the Boltzmann constant. A
parametrization of the UHECR proton's mean-free-path, using delta
function approximation of the $p\gamma$ cross section, is given by
\begin{equation}
\lambda_p \approx 245.76 
\left( \frac{E_p}{100~{\rm EeV}} \right)^{-3}
\exp\left[ 0.42 \left( \frac{E_p}{100~{\rm EeV}} \right) - 1 \right]
~{\rm Mpc},
\label{uhecr_mfp}
\end{equation}
which reproduces results from numerical calculations with accurate
treatment \cite{Stanev:2000fb} within $\sim 10\%$ in the $\sim
60$--200 EeV range. For reference, $\lambda_p = 539$, 247, 138 and 26
Mpc at $E_p = 60$, 80, 100 and 200 EeV, respectively. We search for
astrophysical sources within redshift $z = 0.06$, which corresponds
to a luminosity distance $d_L = 270.4$ Mpc in $\Lambda$CDM cosmology
with $H_0 = 69.6$ km~s$^{-1}$~Mpc$^{-1}$, $\Omega_M = 0.286$ and
$\Omega_{\Lambda} = 0.714$ \cite{Wright:2006up}. The proper distance,
$d_p = d_L/(1+z)^2 = 241$ Mpc, is similar to $\lambda_p$ at 80 EeV.

We have used the {\it Swift}-BAT 70 month X-ray source catalog
\cite{Baumgartner:2014} to search for astrophysical sources which are
correlated with UHECR and cosmic neutrino events.  In 70 months of
observations, the catalog includes 1210 objects of which 503 objects
are within redshift $\le 0.06$. Out of these 503 X-ray selected
objects at least 18 are simultaneously correlated with the neutrino
events and UHECRs above 100 EeV, see Table \ref{tab:iv}.  The
correlated X-ray sources are all Seyfert (Sy) galaxies except ABELL
2319 which is a galaxy cluster (GC). The X-ray luminosity of these
sources vary between $L_X \approx 10^{43}$--$10^{45}$ erg~s$^{-1}$,
with Cygnus A the most luminous of all.  Note that the PAO
collaboration has also found an anisotropy at $\sim 98.6\%$ CL in
UHECRs with energy $\ge 58$ EeV and within $\sim 18^\circ$ circles
around the AGNs in {\em Swift}-BAT catalog \cite{Baumgartner:2014} at
distance $\le 130$ Mpc and X-ray luminosity $L_X \gtrsim 10^{44}$
erg~s$^{-1}$ \cite{PierreAuger:2014yba}. Our list in Table
\ref{tab:iv} includes NGC 1142 which is also one of the five sources
that dominantly contribute to the anisotropy found in the PAO data
\cite{PierreAuger:2014yba}.

In another correlated source search we have used bright extragalactic
radio sources with flux density $\ge 1$ Jy at 5 MHz from the K\"uhr
catalog \cite{Kuhr:1981}. It has 61 sources within known redshift $\le
0.06$.  Only 3 sources from this catalog are correlated simultaneously
with IceCube neutrinos and UHECRs above 100 EeV, see Table
\ref{tab:iv}.  Two of these sources are Seyfert galaxies and the third
one is a galaxy cluster. There are two common sources, that are
correlated with both neutrinos and UHECRs, between the {\it Swift}-BAT
and K\"uhr catalogs. These are NGC 1068 and PKS 2331-240. Both of them
are Seyfert galaxies.

We have also searched the first AGN catalog (1LAC) published by the
{\it Fermi}-LAT collaboration \cite{1LAC} for possible correlations
with neutrino and UHECR arrival directions but did not find any.

It is interesting note that the cosmic neutrino events (nos.\ 2, 12,
14, 15 and 36) which are strongly correlated with the Fermi
Bubbles\footnote{The centers of the error circles within the Fermi
  bubbles' contours} \cite{Lunardini:2013gva, Lunardini:2014wza},
except for event no.\ 2, do not appear in Table \ref{tab:iv}. This
could be a hint to possible extragalactic \cite{Stecker:2013fxa} and
Galactic \cite{Razzaque:2013uoa} components in the neutrino event
data.

\begin{table*}[tbp]\centering
\small
\ra{1.3}
\begin{tabular}{@{}rcrrrcrrr@{}}\toprule
\parbox[c]{1.5cm}{\raggedleft {Neutrino Event $\#$}} & 
\multicolumn{3}{c}{UHECR} & 
\multicolumn{3}{c}{{\it Swift} X-ray Source Catalog \cite{Baumgartner:2014}} \\
  & RA & Dec & Experiment & Name & $z$ & Type \\ \midrule
1 & 45.6 & $-1.7$ & PAO & NGC 1142 & 0.0289 & Sy2 \\
  &    &        &     & NGC 1194 & 0.0136 & Sy1 \\
  &    &        &     & MCG +00-09-042 & 0.0238 & Sy2 \\
  &    &        &     & NGC 1068  & 0.0038 & Sy2\\
11 & 150.1 & $-10.3$ & PAO   & 2MASX J10084862-0954510  & 0.0573 & Sy1.8 \\
17 & 241.5 & 23      & AGASA &  2MASX J16311554+2352577 & 0.0590 & Sy2 \\
29, 34 & 295.6 & 43.52 & TA  & 2MASX J19471938+4449425 & 0.0539 & Sy2 \\
      &       &       &     & ABELL 2319 & 0.0557 & GC \\
      &       &       &     & Cygnus A   & 0.0561 & Sy2 \\
21 & 352.6 & $-20.2$ & PAO & PKS 2331-240 & 0.0477 & Sy2 \\
2, 24, 25 & 294.5 & $-5.8$ & AGASA & 2MASX J19373299-0613046 & 0.0103 & Sy1.5 \\
34 & 340.6 & 12   & PAO   & MCG +01-57-016 & 0.0250 & Sy1.8 \\
   &       &      &       & MCG +02-57-002 & 0.0290 & Sy1.5 \\
   &       &      &       & UGC 12237      & 0.0283 & Sy2 \\
   & 349.0 & 12.3 & AGASA & NGC 7479       & 0.0079 & Sy2/Liner \\
   &       &      &       & 2MASX J23272195+1524375 & 0.0457 & Sy1 \\
   &       &      &       & NGC 7469                & 0.0163 & Sy1.2 \\
   & 352.6 & $-20.2$ & Haverah Park & NGC 7679 & 0.0171 & Sy2 \\ \midrule \\
\parbox[c]{1.5cm}{\raggedleft {Neutrino Event $\#$}} &
\multicolumn{3}{c}{UHECR} & \multicolumn{3}{c}
{K\"uhr Radio Source Catalog \cite{Kuhr:1981} }\\ 
   & RA & Dec & Experiment &  Name & $z$ & Type \\ \midrule
1  & 45.6 & $-1.7$ & PAO &  NGC 1068 & 0.0038 & Sy2 \\
21 & 352.6 & $-20.8$ & PAO & PKS 2331-240 & 0.0477 & Sy2 \\
34 & 340.6 & 12 & PAO & NGC 7385 & 0.0255 & GC \\
\bottomrule
\end{tabular}
\caption{\label{tab:iv}Sources correlated with UHECRs and neutrino events simultaneously.}
\end{table*}

\subsection{Neutrino and UHECR luminosities for correlated events}

After searching for astrophysical sources correlated with both IceCube
cosmic neutrino events and UHECRs, we calculate their corresponding
fluxes required to produce observed events. First, we describe our
point-source neutrino flux calculation method. We assume a power-law
flux of the following form
\begin{equation}
J_{\nu_\alpha} (E_{\nu})= A_{\nu_\alpha} 
\left( \frac{E_{\nu}}{100~{\rm TeV}} \right)^{-\kappa},
\label{nu_flux}
\end{equation}
which is the same for all 3 flavors: $\alpha=e,\mu,\tau$. We estimate
the normalization factor from the number of neutrinos events $N_{\nu}$
of any flavor \footnote{Here we tacitly assume that flavor
  identification is not efficient.} as
\begin{equation}
A_{\nu_\alpha} = \frac{1}{3} \frac{N_{\nu}}
{T \sum_{\alpha} \int_{E_{\nu 1}}^{E_{\nu 2}} 
dE_\nu \,A_{{\rm eff}, \alpha}(E_\nu) 
\left( \frac{E_{\nu}}{100~{\rm TeV}} \right)^{-\kappa} },
\label{nu_flux_normalization}
\end{equation}
where $T$ is IceCube lifetime and $A_{{\rm eff}, \alpha}$ is effective
area for different flavors. We use $T = 988$ days for IceCube 3-year
data release \cite{Aartsen:2014gkd} as in our correlation analysis and
the following parametrization, correct within $\sim 10\%$ uncertainty,
of the effective areas \cite{Aartsen:2013jdh}

\begin{eqnarray}
 A_{{\rm eff}, e} &=& \begin{cases} \left[ 1.26 \times 10^{-5}\, 
(E_{\nu}/{\rm TeV})^{2.64} - 0.017 \right] ~{\rm m}^2 ,  
\hspace{1.0 cm} 25~\text{TeV} \, < E_{\nu} < \,100~\text{TeV} \cr
\left[ 0.459\, (E_{\nu}/{\rm TeV})^{0.5} 
- 1.109 \right] ~{\rm m}^2 ,  \hspace{2.2 cm} E_{\nu} > \,100~\text{TeV} 
\end{cases}
\\ \nonumber
A_{{\rm eff}, \mu} &=& \begin{cases} \left[ 3.6 \times 10^{-6}\, 
(E_{\nu}/{\rm TeV})^{2.64} - 0.0142 \right] ~{\rm m}^2 ,  
\hspace{1.0 cm} 25~\text{TeV} \, < E_{\nu} < \,100~\text{TeV} \cr
\left[ 0.389\, (E_{\nu}/{\rm TeV})^{0.5} - 1.868 \right] ~{\rm m}^2 ,  
\hspace{2.2 cm} E_{\nu} > \,100~\text{TeV} 
\end{cases}
\\ \nonumber
A_{{\rm eff}, \tau} &=& \begin{cases} \left[ 7.267 \times 10^{-6}\, 
(E_{\nu}/{\rm TeV})^{2.64} - 0.0175 \right] ~{\rm m}^2 ,  
\hspace{0.7 cm} 25~\text{TeV} \, < E_{\nu} < \,100~\text{TeV} \cr
\left[ 0.5069\, (E_{\nu}/{\rm TeV})^{0.5} - 3.092 \right] ~{\rm m}^2 ,  
\hspace{2.1 cm} E_{\nu} > \,100~\text{TeV}.
\end{cases}
\label{nu_eff_areas}
\end{eqnarray}

For the limits of the integral in Eq.~(\ref{nu_flux_normalization}) we
set $E_{\nu 1} = 25$ TeV and $E_{\nu 2} = 2.2$ PeV, reflecting
uncertainty in energy estimate reported by the IceCube collaboration
\cite{Aartsen:2014gkd}.

We use neutrino flux to calculate neutrino luminosity of the
corresponding source as
\begin{equation}
L_\nu = 4\pi d_L^2 \sum_\alpha 
\int_{E_{\nu 1}(1+z)}^{E_{\nu 2}(1+z)} dE_\nu \,
E_\nu J_{\nu,\alpha} (E_\nu).
\label{nu_luminosity}
\end{equation}
These luminosities are listed in Table \ref{tab:v}. We use two values
of $\kappa$, the choice $\kappa = 2.3$ is motivated by fit to IceCube
data assuming an isotropic distribution of events
\cite{Anchordoqui:2013qsi, Aartsen:2014gkd} and the choice $\kappa =
2.1$ is motivated by the cosmic-ray spectrum expected from Fermi
acceleration mechanisms. Note that $N_\nu = 3$ for 2MASX
J19373299-0613046, $N_\nu = 2$ for 2MASX J19471938+4449425 and $N_\nu
= 1$ for all other sources. Neutrino luminosities listed in Table
\ref{tab:v} are within a factor 5 of the corresponding X-ray
luminosities of the sources. For the radio sources, neutrino
luminosity far exceeds the corresponding radio luminosity.

\begin{table*}[tbp]\centering
\ra{1.4}
\begin{tabular}{@{}rcrcrrcrr@{}}\toprule
{Source name} & \parbox[c]{2.9cm}
{\raggedleft $L_X \left(10^{44}~\text{erg/s} \right)$}
& \multicolumn{2}{c} {$L_{{\nu}}$
$\left(10^{44}~\text{erg/s}\right) $ }
& \multicolumn{2}{c} {$L_{\text{cr}}$
$\left(10^{44}~\text{erg/s}\right)$} \\
& $/L_R \left( 10^{41}~\text{erg/s} \right)$
& $\kappa=2.1$ & $=2.3$ &  $\kappa=2.1$ & $=2.3$
\\ \midrule
NGC 1142                   & $1.58$/0.012(74 GHz)        & 0.95  & 1.0   & 0.7    & 5.4\\
NGC 1194                   & $0.12$/$0.00012$(1.4 GHz)   & 0.2   & 0.2   & 0.04   & 0.2  \\
MCG +00-09-042             & $0.17$/$0.0043$(1.4 GHz)    & 0.64  & 0.71  & 0.3    & 2.1\\
NGC 1068                   &  $0.031$/0.0034(31.4 GHz)   & 0.016 & 0.017 & 0.001  & 0.007\\
2MASX J10084862-0954510    & $1.04$/$0.0028$(1.4 GHz)    & 3.9   & 4.32  & 44     & 578\\
2MASX J16311554+2352577    & $0.79$/$0.0048$(1.4 GHz)    & 4.1   & 4.6   & 1600   & 22000\\
2MASX J19471938+4449425    & $1.66$/$0.0045$(1.4 GHz)    & 6.8   & 7.6   & 211    & 26000\\
ABELL 2319                 & $1.78$/$0.0046$(1.4 GHz)    & 3.7   & 4.1   & 270    & 3500\\
Cygnus A                   & $11.2$/$314$(14.7 GHz)      & 3.7   & 4.1   & 290    & 3700\\
PKS 2331-240               & $0.81$/$1.32$(31.4 GHz)     & 2.6   & 2.9   & 9.5    & 102\\
2MASX J19373299-0613046    & $0.055$/$0.0012$(1.4 GHz)   & 0.24  & 0.26  & 1.3    & 7.3\\
MCG +01-57-016             & $0.23$/$0.0026$(1.4 GHz)    & 0.71  & 0.78  & 0.5    & 3.6\\
MCG +02-57-002             & $0.25$/$0.00084$(1.4 GHz)   & 0.95  & 1.1   & 1.0    & 7.5\\
UGC 12237                  & $0.23$/$0.0011$(1.4 GHz)    & 0.91  & 1.    & 0.9    & 6.6\\
NGC 7479                   & $0.029$/$0.04$(22 GHz)      & 0.07  & 0.08  & 0.3    & 1.4\\
2MASX J23272195+1524375    & $0.51$/$0.24$(1.4 GHz)      & 2.4   & 2.7   & 280    & 2900\\
NGC 7469                   & $0.4$/$0.0056$(365 MHz)     & 0.3   & 0.3   & 2.2    & 14\\
NGC 7679                   & $0.1$/$0.00033$(1.4 GHz)    & -     & -     & -      & -\\
\midrule
NGC 1068                   &$0.031$/0.0034(31.4 GHz)     & 0.016 & 0.017 & 0.001  & 0.007\\
PKS 2331-240               & $0.81$/$1.32$(31.4 GHz)     & 2.6   & 2.9   & 9.5    & 102\\
NGC 7385                   & - /$0.17$(31.4 GHz)         & 0.7   & 0.8   & 0.5    & 4.0\\
\bottomrule
\end{tabular}
\caption{\label{tab:v}Neutrino (25 TeV--2.2 PeV) and cosmic-ray (500 TeV--180 EeV) luminosities required for the correlated sources in Table \ref{tab:iv} to produce observed data. Also listed are {\it Swift}-BAT X-ray luminosity \cite{Baumgartner:2014} radio luminosity for these sources, with corresponding radio frequencies in parentheses.}
\end{table*}

We also calculate UHECR flux from the observed events by using
exposure for the respective experiments. We use a power-law form for
the observed UHECR flux above 28.8 EeV break energy as
\cite{Abraham:2010mj}
\begin{equation}
J_{\rm uhecr} (E_{\rm cr})= A_{\rm uhecr} 
\left( \frac{E_{\rm cr}}{\rm EeV} \right)^{-4.3} \, ;\, 
E_{\rm cr} \ge 28.8 ~{\rm EeV},
\label{cr_flux}
\end{equation}
and derive the normalization factor as
\begin{equation}
A_{\rm uhecr} = \frac{N_{\rm uhecr}}
{ \frac{\Xi \omega(\delta)}{\Omega} 
\int_{E_{{\rm cr} 1}}^{E_{{\rm cr} 2}} 
dE_{\rm cr} \left( \frac{E_{\rm cr}}{{\rm EeV}} \right)^{-4.3} }.
\label{cr_flux_normalization}
\end{equation}
Here $\Xi$ is the total integrated exposure, as mentioned in
ref.~\cite{Cuoco:2007aa}, $\Omega$ is the solid angle of the detector
and $\omega(\delta)$ is the relative exposure for particular
declination angle $\delta$. For reference, we use for PAO, $\Xi_{\rm
  PAO} = 66,000$~km$^2$~yr~sr and $\Omega_{\rm PAO} = 1.65\pi$~sr
\cite{PierreAuger:2014yba}; for TA, $\Xi_{\rm TA} =
3,690$~km$^2$~yr~sr and $\Omega_{\rm TA} = 0.85\pi$~sr
\cite{Abbasi:2014lda}; for AGASA, $\Xi_{\rm AGASA} =
1,000$~km$^2$~yr~sr and $\Omega_{\rm AGASA} = 0.59\pi$~sr
\cite{Hayashida:2000zr}. We do not use the Haverah Park event that is
correlated with a neutrino event. We calculate $\omega(\delta)$ from
ref.~\cite{Sommers:2000us} but adapt it for different experiments by
using their respective geographical locations and zenith angle ranges.
For the lower and upper limits of integration in
Eq.~(\ref{cr_flux_normalization}), we use $E_{{\rm cr} 1} = 80$ EeV
and $E_{{\rm cr} 2} = 180$ EeV, allowing a $20\%$ uncertainty for the
100 EeV threshold energy used to search correlation and 150 EeV
maximum energy found for a UHECR correlated with a neutrino event and
a source (see Table \ref{tab:ii}).

\begin{figure}[tbp]
  \centering 
  \includegraphics[width=.6\textwidth]{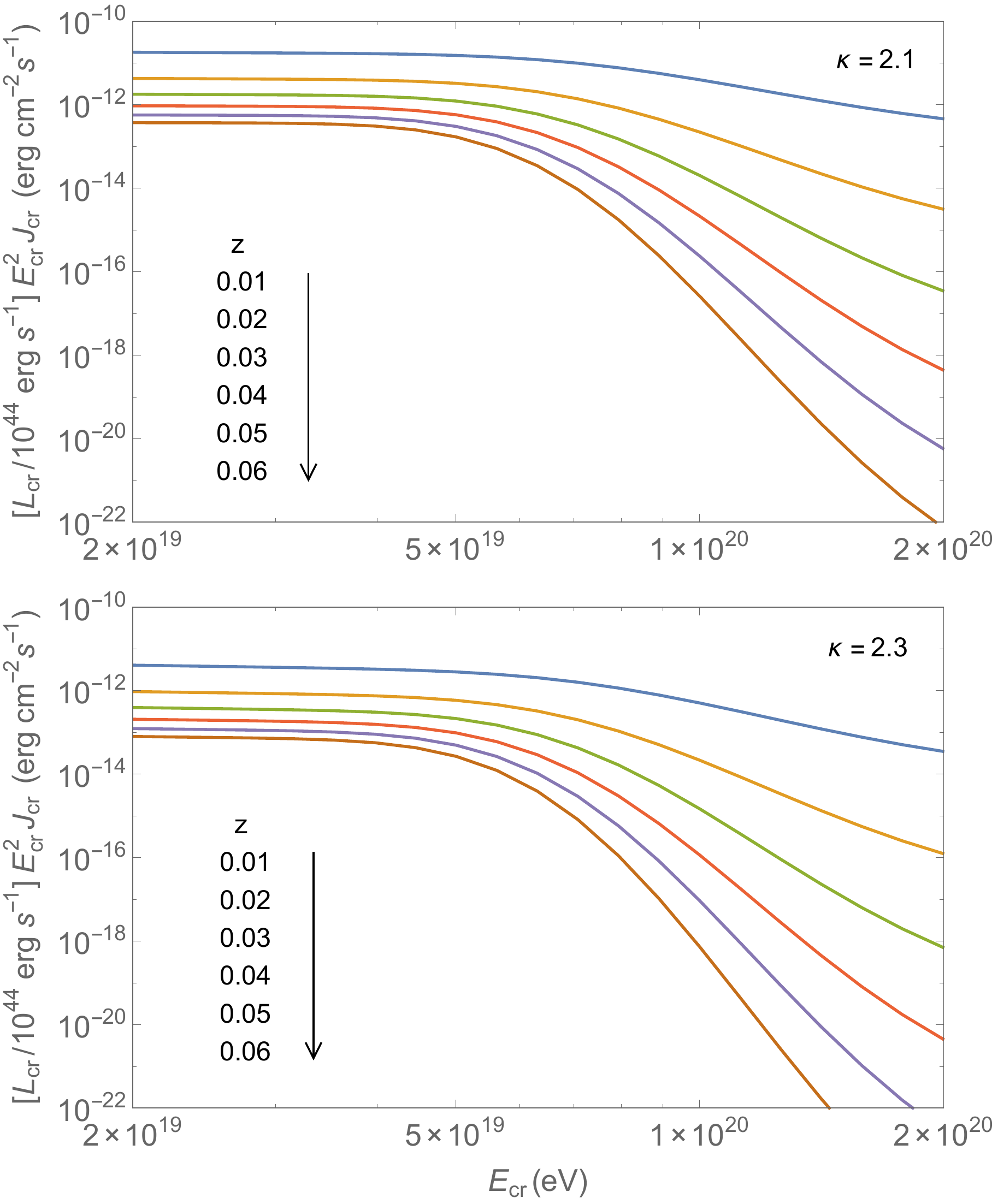}
  \caption{\label{fig:crlum} Expected UHECR flux on the Earth from
    sources at different redshift $0.01\le z\le 0.06$ but with fixed
    luminosity $L_p = 10^{44}$~erg~s$^{-1}$ in the $500$~TeV to 200
    EeV range.}
\end{figure}

In order to calculate cosmic-ray\footnote{We assume they are
  dominantly protons.} luminosity of the sources, first we note that
if cosmic neutrinos detected by IceCube are coming from the same
source that are correlated with UHECRs, then the cosmic-ray flux in
Eq.~(\ref{cr_flux}) needs to be extrapolated down to $\sim 500$ TeV
which is required to produce $\sim 30$ TeV neutrinos. Second, the flux
in Eq.~(\ref{cr_flux}) needs to be corrected for GZK suppression above
$\sim 40$ EeV. Given a cosmic-ray proton luminosity $L_{\rm cr}$
between the generation energies $E^\prime_{{\rm cr} 1} = 500$~TeV and
$E^\prime_{{\rm cr} 2} = 180$~EeV with $\propto E_{\rm cr}^{\prime
  -\kappa}$ spectrum in a source at redshift $z$, the cosmic-ray flux
on the Earth is \cite{Berezinsky:2002nc, Razzaque:2011jc}
\begin{equation}
J_{\rm cr} (E_{\rm cr}) = \frac{L_{\rm cr} (1+z)}{4\pi d_L^2} 
\frac{(\kappa -2)(E^\prime_{{\rm cr}1} E^\prime_{{\rm cr}2})^{\kappa - 2}}
{E_{{\rm cr}2}^{\prime \kappa - 2} - E_{{\rm cr}1}^{\prime \kappa - 2}}
E_{\rm cr}^{\prime -\kappa}
\left( \frac{dE_{\rm cr}^\prime}{dE_{\rm cr}} \right).
\label{cr_flux2}
\end{equation}
The cosmic-ray energy at the source and on the Earth are related
through various energy losses \cite{Berezinsky:2002nc}. Following
ref.~\cite{Razzaque:2011jc} we have plotted cosmic-ray flux in
Fig.~\ref{fig:crlum} using Eq.~(\ref{cr_flux2}) for $L_{\rm cr} =
10^{44}$~erg~s$^{-1}$ and for various redshift in the range $0.01\le
z\le 0.06$. We have also used two different values for $\kappa$ as we
did for neutrino flux calculation.

Figure~\ref{fig:crlum} provides a map to estimate cosmic-ray
luminosity of the sources listed in Table \ref{tab:iv} which are
correlated with UHECR and neutrino events. The UHECR flux in
Eq.~(\ref{cr_flux}), calculated from data, corresponds to a point in
Fig.~\ref{fig:crlum} at $E_{\rm cr} \approx 80$ EeV. We estimate the
source luminosity $L_{\rm cr}$ by equating this flux to the expected
flux in Eq.~(\ref{cr_flux2}) at 80 EeV for the redshift of a given
source.  These luminosities are listed in Table \ref{tab:v}. Note that
except for NGC 1068 ($z=0.0038$), NGC 1194 ($z=0.0136$) and MCG
+00-09-042 ($z=0.0238$) the cosmic-ray luminosity with $\kappa = 2.1$
is comparable or higher than the neutrino luminosity for all sources.
The cosmic-ray luminosity exceeds the X-ray or radio luminosities for
all sources except for NGC 1142 and NGC 1194, in case $\kappa = 2.1$.


\section{Discussion and Outlook}
\label{discussion}

We have investigated whether the arrival directions of cosmic
neutrinos, detected by IceCube \cite{Aartsen:2014gkd}, with energy
$\sim 30$ TeV--2 PeV are correlated with the arrival directions of
UHECRs with energy $\gtrsim 100$ EeV. In order to test correlation we
have used an invariant statistic, called the minimum $\delta \chi^2$
\cite{Virmani:2002xk}, which is constructed from the angle between two
unit vectors corresponding to the directions of the neutrino events
and UHECRs, and weighted by the angular resolutions of the neutrino
events. We have evaluated the significance of any correlation by using
Monte Carlo simulations of randomly generated UHECR directions and
comparing with data. We found that IceCube cosmic neutrinos are
correlated with UHECRs with energy $\ge 100$ EeV with significance at
$90\%$ CL. The significance, however, decreases with decreasing energy
of UHECRs, { leaving no correlation at an energy threshold
  of $60$ EeV}.
To take into account trial factor, since we searched for correlation
with $N_{\rm trial} = 3$ UHECR energy thresholds, we calculate
post-trial $p$-value as $p_{\rm post-trial} = 1-(1-p_{\rm
  signal})^{1/N_{\rm trial}} = 0.27$, with $p_{\rm signal} = 0.1$ that
we found in data.

We have searched for astrophysical sources in the {\it Swift}-BAT
X-ray catalog \cite{Baumgartner:2014}, the K\"uhr radio source catalog
\cite{Kuhr:1981} and {\it Fermi}-LAT 1LAC AGN catalog \cite{1LAC}
within $3 ^{\circ}$ error circles of the $\ge 100$ EeV UHECRs which
are correlated with cosmic neutrino events, assuming the UHECRs are
protons. We made a cut in redshift, $z \le 0.06$, while searching for
sources in the catalogs. This corresponds to a proper distance of 241
Mpc, similar to the mean-free-path of an 80 EeV proton in the CMB. The
choice of $3^{\circ}$ error circle is motivated by deflection of UHECR
protons in the intergalactic and Galactic magnetic fields. We found
that 18 sources from the {\it Swift}-BAT X-ray catalog and 3 sources
from the K\"uhr radio source catalog are within $3^{\circ}$ error
circles of the UHECRs that are correlated with cosmic neutrinos.
Except for ABELL 2319 and NGC 7385 which are galaxy clusters, the rest
of the sources are Seyfert galaxies with Cygnus A being the most well
known.  Our finding is consistent with that of the PAO collaboration
who found significant correlation between UHECR arrival directions and
Seyfert galaxies in the {\it Swift}-BAT X-ray catalog
\cite{PierreAuger:2014yba}.  We did not find any source from the {\it
  Fermi}-LAT 1LAC AGN catalog fitting our search criteria.

Estimates of the neutrino and UHECR fluxes for the correlated events
were used to calculate corresponding 25 TeV--2.2 PeV neutrino
luminosity and 500 TeV--180 EeV cosmic-ray luminosity under the
hypothesis that both originated from the sources we found in the {\it
  Swift}-BAT and K\"uhr catalogs. The neutrino luminosities are of the
same order as the X-ray luminosities of the sources. The cosmic-ray
luminosities, depending on the source spectrum, are comparable or
higher than both the neutrino and X-ray luminosities.  Comparison
between the nonthermal X-ray luminosity with the cosmic-ray or
neutrino luminosity gives a possibility that the energy in X-ray
producing electrons can be compared to that of cosmic-ray protons,
both accelerated at the sources.

Acceleration of UHECRs near the central black holes of AGNs was
proposed over 20 years ago~\cite{ste, pro}.  Interactions of these
UHECRs with UV and X-ray photons could produce high-energy neutrinos
\cite{ste, Szabo:1994qx}. Seyfert galaxies are radio-quite AGNs and do
not have strong jets, although parsec scale jets in them have been
observed in the last decade \cite{mun,gal1,mil}. Collisions between
blobs in this jet and formation of shocks may lead to acceleration of
protons to an energy at least up to $10^{18}$ eV, with subsequent
photomeson interactions producing high-energy neutrinos
\cite{AlvarezMuniz:2004uz}. Acceleration of heavy nuclei and
subsequent gamma-ray and neutrino production in radio-quite AGNs have
also been discussed \cite{Pe'er:2009rc}. Predictions have also been
made for GeV--TeV gamma-ray emission from UHECR interactions in Cygnus
A \cite{Atoyan:2008uy}, which is also a powerful radio galaxy (3C
405).

If a fraction $\eta_X$ of the X-ray luminosity, $L_X = 10^{44}
L_{44}$~erg~s$^{-1}$, of the Seyfert galaxies is nonthermal then one
can estimate the energy density in magnetic field in the X-ray
emitting region as $B^2/8\pi = \eta_X L_X /4\pi R^2 c$. Using $R
\approx 10^{14} R_{14}$ cm, 3 times the Schwarzschild radius of a
black hole of mass $M_{\rm bh} = 10^8 M_\odot$, the magnetic field is
$B = 10^3 (\eta_X L_{44})^{1/2} R_{14}^{-1}$~G.  Assuming protons are
accelerated in the same region, their maximum energy can be $E_{\rm
  max} = eBR = 2.4\times 10^{19}(\eta_X L_{44})^{1/2}$~eV.  This is
problematic for the X-ray luminosities of Seyfert galaxies in Table
\ref{tab:v}, which are correlated with $\ge 100$ EeV cosmic rays, and
the magnetic energy density must exceed the nonthermal X-ray energy
density by a factor $\gtrsim 10$ for proton acceleration to $\sim
10^{20}$ eV. This additional energy could be accommodated if a
sizable fraction of the Eddington luminosity, $L_{\rm Edd} =
1.3\times 10^{46} (M_{\rm bh}/10^8 M_\odot)$~erg~s$^{-1}$, could be
converted to magnetic energy. The required cosmic-ray luminosities
($\kappa = 2.1$) in Table \ref{tab:v} for Seyfert galaxies are above
the Eddington luminosity for 2MASX J16311554+2352577, 2MASX
J19471938+4449425, 2MASX J23272195+1524375.  In case of Cygnus A,
$M_{\rm bh}$ and $L_{\rm Edd}$ are an order of magnitude larger. The
$\kappa=2.3$ cosmic-ray luminosities are more problematic.  The
opacity for photomeson ($p\gamma$) interactions with $\epsilon_X =
1$~keV X-ray photons and the subsequent $\gtrsim 15$ TeV neutrino
production opacity is $\tau_{p\gamma} \approx 1\,
L_{44}R_{14}^{-1}(\epsilon_X/1~{\rm keV})^{-1}$.

A hint of correlation that we found between the IceCube cosmic
neutrino events and UHECRs with energy $\ge 100$ EeV should be
investigated further by the experimental collaborations. Establishing
a concrete correlation will be a ground-breaking discovery. Also a
future extension of IceCube to increase its sensitivity in the $> 1$
PeV range will be very useful to probe the cosmic neutrino spectrum
and if there is a cutoff in the spectrum. A cutoff in the spectrum is
not natural at the PeV scale if the same sources produce $\ge 100$ EeV
cosmic rays and neutrinos. Whether the weak AGNs, which are plentiful
in the nearby universe, are the sources of UHECRs and neutrinos or not
is a question that will continued to be debated and investigated in
the years to come.

\acknowledgments

We thank Paul Sommers for useful comments. We also
  thank an anonymous referee for helpful suggestions to improve this
  work. This work was supported in part by the National Research
Foundation (South Africa) grants nos.\ 87823 (CPRR) and 91802 (Blue
Skies). This research has made use of the VizieR catalog access tool,
CDS, Strasbourg, France. The original description of the VizieR
service was published in A\&AS 143, 23.



\end{document}